\documentclass[prl,twocolumn,byrevtex,floatfix,superscriptaddress,longbibliography]{revtex4}
\usepackage{gensymb}
\usepackage{amsmath}    
\usepackage{amsfonts}

\usepackage{graphicx}   
\usepackage{verbatim}   
\usepackage{color}      
\usepackage{subfigure}  
\usepackage{hyperref}   
\setcitestyle{super}

\begin{document}

\title{\boldmath Fast non-volatile electric control of antiferromagnetic states \unboldmath}

\author{S. Ghara\footnote{email: somnath.ghara@physik.uni-augsburg.de}}
\affiliation{Experimental Physics V, Center for Electronic Correlations and Magnetism, University of Augsburg, 86159 Augsburg, Germany}
\author{M. Winkler}
\affiliation{Experimental Physics V, Center for Electronic Correlations and Magnetism, University of Augsburg, 86159 Augsburg, Germany}
\author{K. Geirhos}
\affiliation{Experimental Physics V, Center for Electronic Correlations and Magnetism, University of Augsburg, 86159 Augsburg, Germany}
\author{L. Prodan}
\affiliation{Experimental Physics V, Center for Electronic Correlations and Magnetism, University of Augsburg, 86159 Augsburg, Germany}
\author{V. Tsurkan}
\affiliation{Experimental Physics V, Center for Electronic Correlations and Magnetism, University of Augsburg, 86159 Augsburg, Germany}
\affiliation{Institute of Applied Physics, Moldova}
\author{S. Krohns}
\affiliation{Experimental Physics V, Center for Electronic Correlations and Magnetism, University of Augsburg, 86159 Augsburg, Germany}
\author{I. K\'ezsm\'arki}
\affiliation{Experimental Physics V, Center for Electronic Correlations and Magnetism, University of Augsburg, 86159 Augsburg, Germany}


\begin{abstract}
\textbf{Electrical manipulation of antiferromagnetic states, a cornerstone of antiferromagnetic spintronics, is a great challenge, requiring novel material platforms. Here we report the full control over antiferromagnetic states by voltage pulses in the insulating Co$_3$O$_4$ spinel. We show that the strong linear magnetoelectric effect emerging in its antiferromagnetic state is fully governed by the orientation of the N\'eel vector. As a unique feature of Co$_3$O$_4$, the magnetoelectric energy can easily overcome the weak magnetocrystalline anisotropy, thus, the N\'eel vector can be manipulated on demand, either rotated smoothly or reversed suddenly, by combined electric and magnetic fields. We succeed with switching between antiferromagnetic states of opposite N\'eel vectors by voltage pulses within a few microsecond in macroscopic volumes. These observations render quasi-cubic antiferromagnets, like Co$_3$O$_4$, an ideal platform for the ultrafast (pico- to nanosecond) manipulation of microscopic antiferromagnetic domains and may pave the way for the realization of antiferromagnetic spintronic devices.}
\end{abstract}

\maketitle

Ferromagnets still represent an important material platform for magnetic storage and spintronic devices~\cite{vzutic2004,dietl2014}. However, they set restrictions on further miniaturization and speed-up of these devices due to several drawbacks, such as the cross-talking between the adjacent units via their stray fields, high sensitivity to external magnetic fields and relatively slow spin dynamics~\cite{jungwirth2016,baltz2018}. 

Though antiferromagnets are free from these limitations, their applicability has not been put into consideration for long. Even Louis N\'eel, who discovered them, stated in his Nobel lecture that antiferromagnets had no practical importance~\cite{neel1971}. Contradicting to his prediction, a plethora of recent studies imply that antiferromagnets can be viable replacements for ferromagnetic components in spintronic devices~\cite{jungwirth2016,baltz2018,jungwirth2018natphys,jungfleisch2018}. In fact, emergent phenomena associated with antiferromagnets, such as anisotropic magnetoresistance~\cite{fina2014,wang2014}, spin Hall effect~\cite{zhang2014,kimata2019,chen2021}, spin Seebeck effects~\cite{seki2015,wu2016,rezende2016}, N\'eel spin-orbit torque~\cite{zelezny2014,wadley2016}, ultrafast THz manipulation of antiferromagnetic (AFM) spin waves~\cite{kampfrath2011} and dynamics of AFM skyrmions~\cite{gao2020,legrand2020,zhang2016,woo2018}, open up new opportunities in spintronics. In fact, AFM spintronics, which is an impressively growing field at the frontiers of magnetism and electronics, is foreseen to trigger a change of paradigm in information technology~\cite{jungwirth2016,baltz2018,jungwirth2018natphys,duine2018}.

In the last years, the electrical manipulation of AFM domains and the orientation of the AFM N\'eel vector ($\mathbf{L}$) have attracted much attention, as a possible route to write-in and read-out information in AFM devices~\cite{jungwirth2016,baltz2018,jungfleisch2018,song2018,leo2018,parthasarathy2019,baldrati2020,yan2020}. The spin-orbit coupling plays a key role in controlling the orientation of $\mathbf{L}$ in itinerant and insulating antiferromagnets via electric currents and voltages, respectively~\cite{thole2020}. In non-centrosymmetric AFM metals, like CuMnAs~\cite{wadley2016}, Mn$_2$Au~\cite{bodnar2018} and Fe$_{1/3}$NbS$_2$~\cite{nair2020}, the current induced rotation of $\mathbf{L}$ is governed by the spin-orbit torque and has led to the realization of the first prototype of AFM random access memories. In non-centrosymmetric AFM insulators, such as Cr$_2$O$_3$~\cite{borisov2005,iyama2013}, LiCoPO$_4$~\cite{kocsis2018,van2007,zimmermann2014} and MnTiO$_3$~\cite{sato2020,mufti2011}, the linear magnetoelectric (ME) effect, that is another manifestation of the spin-orbit coupling, can be exploited for the electric field-driven switching of AFM domain states, as a working principle for AFM-ME random access memories~\cite{kosub2017}. While the voltage-driven switching of $\mathbf{L}$ has an advantage over current-driven switching in terms of heat dissipation, especially for devices with high density of information and/or active elements, such in situ voltage-driven switching has been observed only in a few compounds so far, and in all cases only in the vicinity of $T_N$~\cite{he2010,iyama2013,kosub2017,sato2020}.

In this proof-of-concept study, we demonstrate the in situ isothermal switching of $\mathbf{L}$ driven by an electric field deep in the AFM state of insulating Co$_3$O$_4$. This compound belongs to the family of only $A$-site magnetic spinels, where Co$^{2+}$ ions, occupying the diamond lattice formed by the $A$ sites, have spins S=3/2 in tetrahedral oxygen coordination \cite{roth1964,cova2019,sunda2021}. The pyrochlore lattice of the $B$ sites is occupied by non-magnetic Co$^{3+}$ ions. While Co$_3$O$_4$ has a centrosymmetric crystal structure (space group $Fd\bar{3}m$) in the paramagnetic state, the Co$^{2+}$ ions have no inversion symmetry (site symmetry $\bar{4}3m$)~\cite{saha2016} and the onset of N\'eel-type AFM order at $T_N$=30\,K breaks the global inversion symmetry of the lattice, making Co$_3$O$_4$ a non-centrosymmetric insulating antiferromagnet. Earlier reports showed that Co$_3$O$_4$ undergoes a transition to a simple textbook-like collinear AFM state at $T_N$~\cite{roth1964,saha2016}, in strong contrast to the other highly frustrated $A$-site spinels, such as $M$Sc$_2$S$_4$ ($M$=Mn, Fe)~\cite{gao2017,fritsch2004}. However, the direction of the magnetic easy-axis, which can be either along the $\langle$111$\rangle$-type body-diagonals or along the $\langle$001$\rangle$-type main cubic axes, has not been resolved yet due to the scarce of single crystals. 

This compound, along with a few other $A$-site magnetic spinels, such as CoAl$_2$O$_4$ and Mn$B_2$O$_4$ ($B$=Al, Ga), has recently been reported to exhibit a linear ME effect~\cite{ter2014,saha2016,ghara2017,de2021}. Due to the non-centrosymmetric tetrahedral environment ($\bar{4}3m$) of the magnetic Co$^{2+}$ ions, the ME free energy has the following form~\cite{ter2014}:
\begin{multline}
F_{ME}=L_x(M_yP_z+M_zP_y)+L_y(M_zP_x+M_xP_z)\\
+L_z(M_xP_y+M_yP_x).
\end{multline}
Here, the $x$, $y$ and $z$ axes are the main cubic axes, whereas $M$, $L$ and $P$ stands for the magnetization, the AFM N\'eel vector and the electric polarization, respectively. For weak magnetic fields, when $M\ll L$ and the magnitude of $\mathbf{L}$ is nearly field independent, Eq.~(1) yields a linear ME susceptibility ($\hat{\alpha}$), which solely depends on $\mathbf{L}$ according to,
\begin{equation}
\hat{\alpha}\sim
\begin{pmatrix} 
0 & L_z & L_y \\
L_z & 0 & L_x\\
L_y & L_x & 0 \\
\end{pmatrix}.
\quad
\end{equation}

This specific form of its ME coupling together with its weak magnetic anisotropy make Co$_3$O$_4$ an ideal benchmarking material to investigate the electrical control of $\mathbf{L}$. For example, when applying an electric field along one of the main cubic axes together with a magnetic field arbitrarily oriented in the plane perpendicular to it, the formation of an AFM mono-domain state with $\mathbf{L}$ perpendicular to both fields is expected, if the ME free energy can overcome the magnetocrystalline anisotropy energy. 

Here, we demonstrate the full control over the orientation of $\mathbf{L}$, achieved by simultaneous electric and magnetic fields in the AFM state of Co$_3$O$_4$. The electrical switching of $\mathbf{L}$ in macroscopic volumes is accomplished only within a few microseconds, pointing to an ultrafast intrinsic switching mechanism for microscopic AFM domains.

\begin{figure}[t!]
\includegraphics[width=3in]{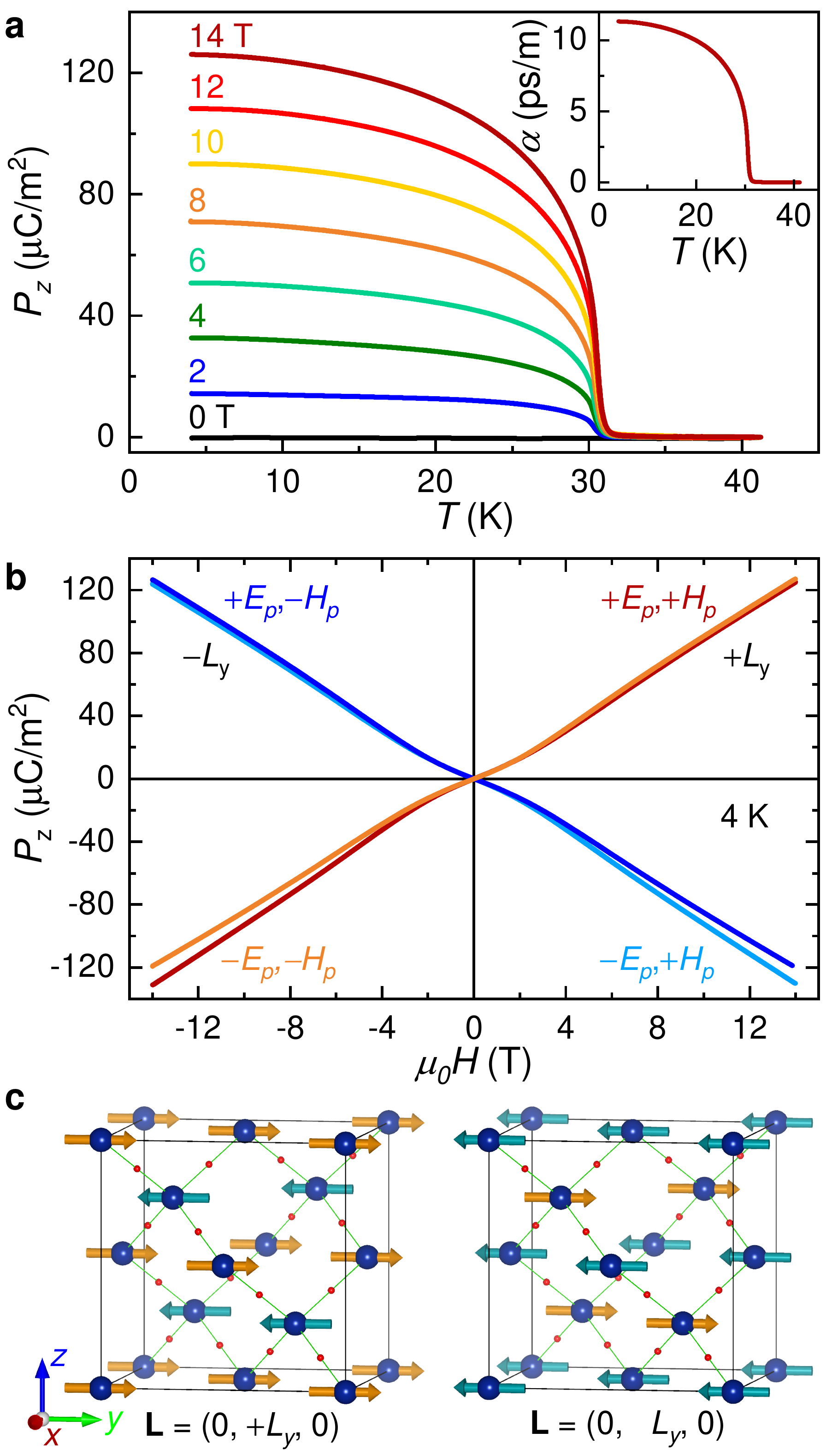}
\caption{\textbf{$\mid$ Realization of two distinct AFM mono-domain states via ME poling.} \textbf{a},~Temperature-dependent electric polarization $P_z$ measured after ME poling with $E_p$=+4.2\,kV/cm ($\parallel z$ axis) and various $\mu_0H_p$ ($\parallel x$ axis). The inset shows the temperature-dependent ME coefficient $\alpha$. \textbf{b},~Magnetic field-dependent electric polarization at 4\,K measured after ME poling with all four sign combinations of the poling fields, $E_p$ = $\pm$4.2\,kV/cm and $\mu_0H_p$ = $\pm$14\,T. In all these cases, the electric polarization was measured with $E_p$ being switched off after ME poling. \textbf{c},~Schematics of $+L_y$ and $-L_y$ AFM mono-domain states in Co$_3$O$_4$. The red dots between the Co$^{2+}$ ions of opposite magnetization indicate the centers of inversion in the paramagnetic cubic state, which are broken by the AFM order. \label{Fig1}}
\end{figure}

\begin{figure*}[t!]
\includegraphics[width=7in]{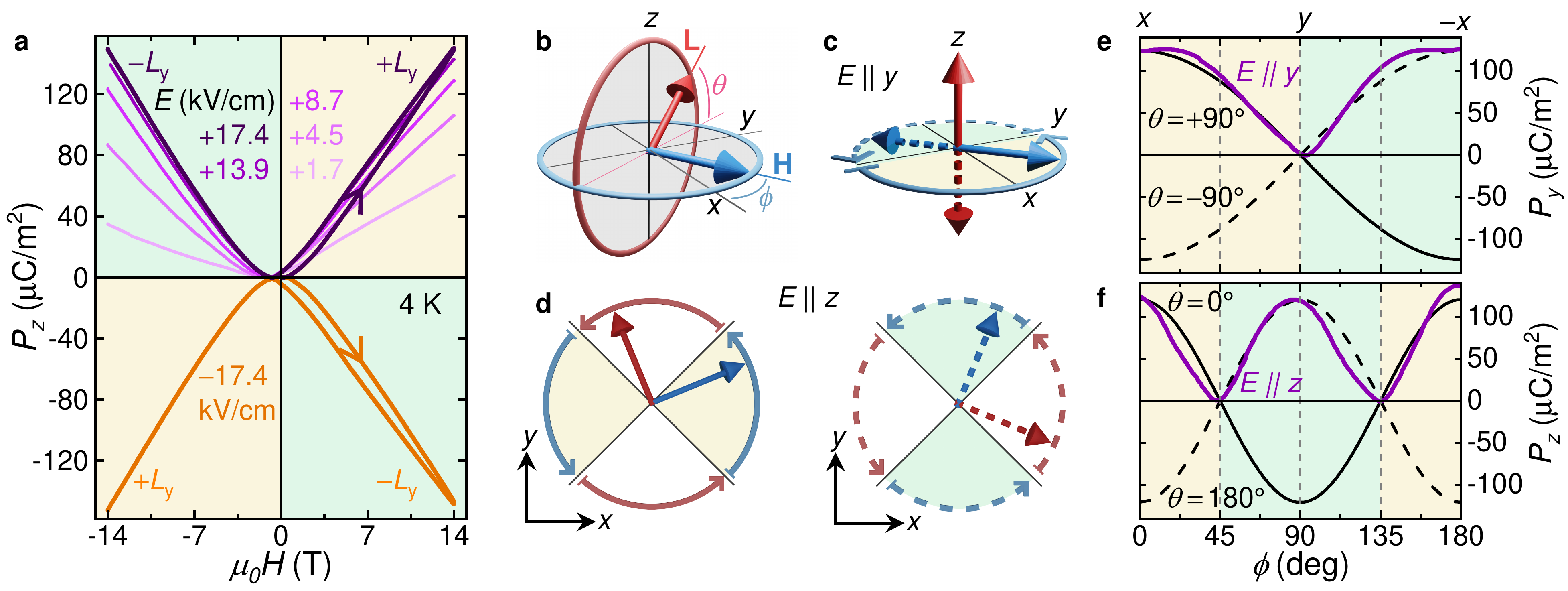}
\caption{\textbf{$\mid$ In situ switching and rotation of $\mathbf{L}$.} \textbf{a},~Magnetic field ($\mathbf{H}$ $\parallel x$ axis) dependence of the electric polarization at 4\,K measured in the presence of electric fields ($\mathbf{E}$ $\parallel z$ axis). \textbf{b},~Sketch showing the orthogonality of the AFM N\'eel vector $\mathbf{L}$ (red arrow) to the magnetic field (blue arrow) and introducing the angles $\phi$ and $\theta$. \textbf{c},~Schematic diagram showing the favoured direction of $\mathbf{L}$, either $+L_z$ or $-L_z$, depending on the orientation of the magnetic field rotated within the $xy$ plane, when the electric field is applied along the $y$ axis. \textbf{d},~Schematics of the in-plane rotation of $\mathbf{L}$ with either preceding (yellow shaded angular range) or following (green shaded angular range) the magnetic field rotating in the $xy$ plane, when the electric field is applied along the $z$ axis. \textbf{e}/\textbf{f},~Dependence of the electric polarization component $P_y$/$P_z$ on the orientation of the magnetic field (14\,T) rotating within the $xy$ plane at 4\,K in the presence of an electric field (+17.4\,kV/cm) applied along $y$/$z$ axis. The purple curves in the two panels are the experimental data. The solid/dashed black curve in panel \textbf{e} indicates the calculated behaviour of $P_y$, according to Eq.~(3), for the $+L_z$/$-L_z$ mono-domain state, and those in panel \textbf{f} show the expected behaviour of $P_z$, according to Eq.~(4), when $\mathbf{L}$ rotates with preceding/following the magnetic field within the $xy$ plane. In all panels, the yellow and green background colours indicate field or angular regions, where $\mathbf{L}$ is preserved (panels \textbf{a}, \textbf{c} \& \textbf{e}) or continuously rotated (panels \textbf{d} \& \textbf{f}). At the borders of these regions $\mathbf{L}$ is switched suddenly.  \label{Fig2}}
\end{figure*}

\vspace{0.2cm}

\textbf{AFM domain selection via ME effect} 

Figure~1 summarizes our results on the linear ME effect obtained after poling with orthogonal electric and magnetic fields, namely $E_p\parallel z$ axis and $H_p\parallel x$ axis. As clear from Fig.~1a, the electric polarization emerges below $T_N$ in proportion to the magnetic field, characteristic to the linear ME effect. The value of $\alpha$ is found to be $\sim$11.3\,ps/m at 4\,K (inset of Fig.~1a), which is four times larger compared to that of the prototypical ME material Cr$_2$O$_3$~\cite{iyama2013}. Note that $\alpha$ stands for the $xz$ component of $\hat{\alpha}$ throughout this work. The linear behaviour of the ME effect is directly evidenced in the magnetic field-dependent electric polarization (Fig.~1b). The sign of $\alpha$, that is the sign of the slope of the $P$-$H$ curve, depends on the relative signs of the poling fields: $\alpha$ is positive/negative when the product $H_pE_p$ is positive/negative. These results, together with the presence/absence of electric polarization for perpendicular/parallel orientation of the electric and magnetic fields (shown in Fig.~2e and discussed below), confirm the specific form of the linear ME effect, as given in Eqs.~(1) \& (2).

For electric and magnetic fields, respectively, applied along the $z$ and the $x$ axes, the ME free energy term has the simple form $L_yM_xP_z$ and, thus, the $+L_y$ AFM mono-domain state is expected to emerge for $H_pE_p>0$ and the $-L_y$ mono-domain state for $H_pE_p<0$ (see Fig.~1c). Indeed, such selection of the $\pm L_y$ mono-domain states for opposite signs of $H_pE_p$ is implied by $P$-$H$ curves in Fig.~1b. Please note that in the present case, $\pm L_y$ AFM states are interchanged not only by time reversal but also by spatial inversion, that is why the sign of $\mathbf{L}$ is linked to the sign of $H_pE_p$. Before measuring the electric polarization at $T$ = 4\,K, shown in Fig.~1b, the poling electric field was switched off. Since the $P$-$H$ curve is linear in the whole magnetic field range, we can conclude that the AFM mono-domain state, established by poling with crossed $E_p$ and $H_p$, was preserved upon the entire magnetic field sweep. We note that magnetic anisotropy with easy axes along the cubic $\langle$111$\rangle$-type directions could prevent the formation of the $\pm L_y$ AFM mono-domain states. However, the magnetization of Co$_3$O$_4$ is found to be independent of the orientation of the magnetic field (see the supplementary figure~1), which indicates the weakness of magnetic anisotropy and the dominance of the external fields in determining the orientation of $\mathbf{L}$ via the ME coupling. 

\begin{figure*}[t!]
\includegraphics[width=7in]{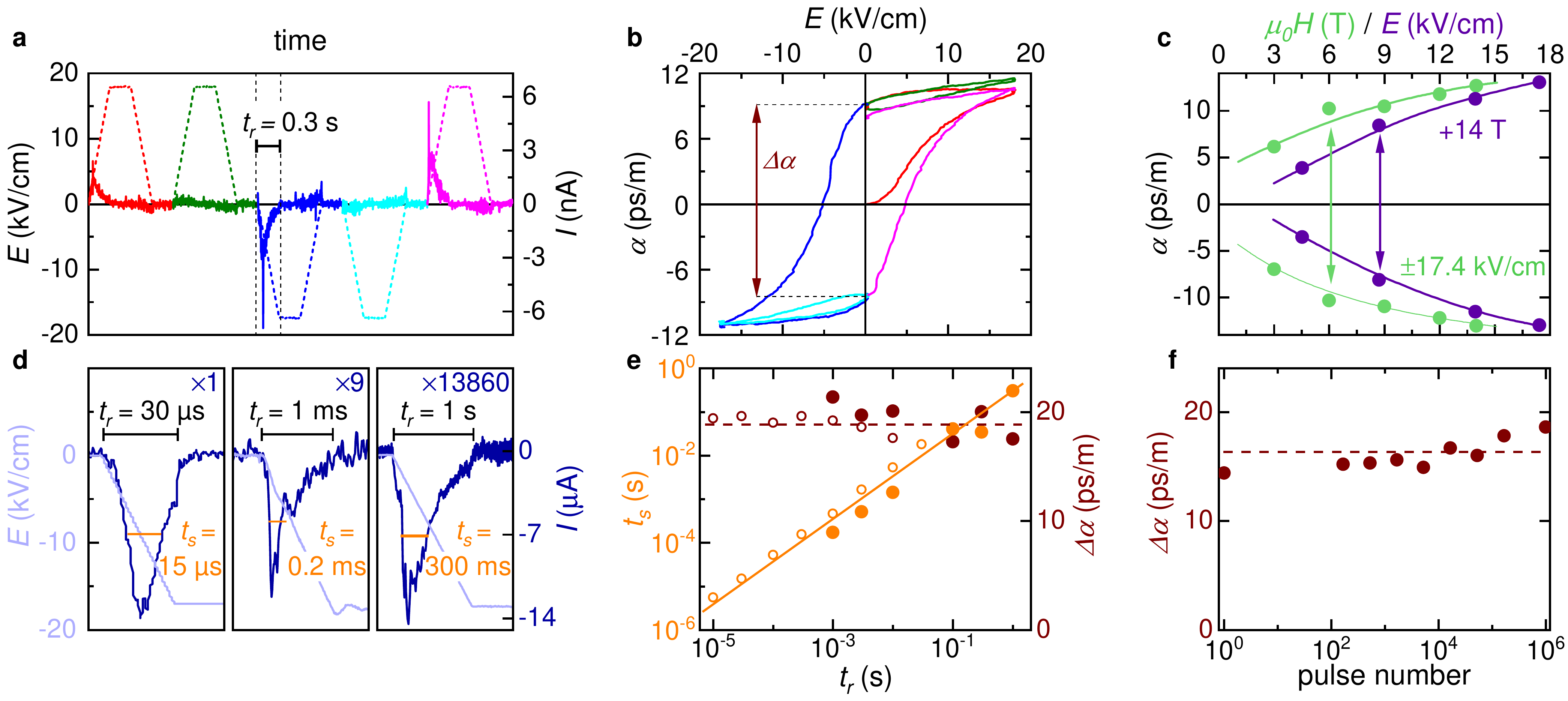}
\caption{\textbf{$\mid$ Voltage pulse-driven switching of $\mathbf{L}$.} \textbf{a},~The $+$, $+$, $-$, $-$, ... electric pulse sequence (dashed line/left scale) for $E \parallel z$ axis and the measured switching current (solid line/right scale) at 4\,K in the presence of +14 T ($\parallel x$ axis). \textbf{b},~ME hysteresis loop at 4\,K in +14\,T. The difference between the $+$ and $-$ remanent value of the ME coefficient, $\Delta\alpha$, is a measure of the switching efficiency. The sequence of pulses is indicated by colours according to panel \textbf{a}. \textbf{c},~Dependence of the ME coefficient $\alpha$ on the strength of the pulsed electric field (purple), when +14\,T is constantly applied,  and on the strength of the static magnetic field (green), while exposing the sample to $\pm$17.4\,kV/cm electric pulses. The $+$/$-$ signs of $\alpha$ corresponds to the $+L_y$/$-L_y$ AFM mono-domain state. \textbf{d},~Time traces of the driving electric pulses (light blue line) for various rise times ($t_r$) and the time traces of the corresponding switching currents (dark blue line). The latter are plotted on a common vertical scale after multiplication with numerical factors, as indicated. The estimated switching time of the AFM domains is indicated by $t_s$. \textbf{e},~Dependence of the switching time $t_s$ (left scale) and the switching efficiency $\Delta\alpha$ (right scale) on the rise time $t_r$ of the driving electric pulse. The data indicated by solid and open symbols were obtained on a thick and a thin crystal, respectively (see Methods). \textbf{f},~Dependence of the switching efficiency on the number of $+$, $-$, ... pulses sequences at 4\,K in +14\,T. The solid and dashed lines in panel \textbf{c}, \textbf{e} and \textbf{f} are guides to the eyes. \label{Fig3}}
\end{figure*}

\vspace{0.2cm}

\textbf{Magnetic control and kinetics of $\mathbf{L}$} 

In the following, we demonstrate that the AFM mono-domain state can be achieved not only by ME poling but can also be created in situ from a multi-domain state by orthogonal electric and magnetic fields, even at temperatures well below $T_N$. In addition to the mono-domain creation, we control the orientation of $\mathbf{L}$ in two ways: We reverse it via the reversal of either the electric or the magnetic field, and rotate it by rotating the magnetic field, while the electric field is kept constant.  

For these studies, the sample was cooled to 4\,K without any poling fields. In the absence of electric field, no magnetically induced polarization was observed, indicating that the net $\mathbf{L}$ averages to zero in the multi-domain state. When starting from the virgin multi-domain state and applying a constant electric field of $E$=$+$17.4\,kV/cm along the $z$ axis, the sample is set to the $+L_y$ mono-domain state by magnetic fields exceeding 2 - 3\,T, applied along the $x$ axis. This is shown in Fig.~2a, where the initial parabolic increase of the $P$-$H$ curve turns to linear above this field range, implying no further changes in domain population. Indeed, in $+$14\,T the electric polarization reaches the same value as recorded after ME poling (cf. Figs.~1a and 1b), also evidencing that the $+L_y$ mono-domain state has been achieved. Intriguingly, when sweeping the magnetic field back with keeping the constant electric field on, the slope of the $P$-$H$ curve changes sign for negative magnetic fields, indicating the switching to the $-L_y$ mono-domain state. This is opposite to the case with no electric field applied, when the poled $+L_y$ mono-domain state is preserved upon the sign change of the magnetic field (Fig.~1b). Electric fields as low as $+$1.7\,kV/cm are still efficient in increasing the population of the $+L_y$ state and switching the sign of the excess $+L_y$ component, but the mono-domain state is only reached for $E\ge$ $+$8.7\,kV/cm, as seen in Fig.~2a. The same domain selection and switching processes are also demonstrated in Fig. 2a for the opposite sign of the electric field.

After demonstrating the in situ ME domain control, we investigate the kinetics of $\pm \mathbf{L}$ switching and the rotation of $\mathbf{L}$. This is done by analysing the dependence of electric polarization on the orientation of the magnetic field using Eqs.~(1)-(2). Due to the negligibly small anisotropy energy, $\mathbf{M}$ coaligns with magnetic fields larger than 2-3\,T, irrespective of the field direction, and $\mathbf{L}$ can point along any direction perpendicular to $\mathbf{M}$, as depicted in Fig.~2b. Via the ME free energy, the application of an additional electric field can select an unique $\mathbf{L}$, which determines the electric polarization, probed experimentally, according to Eq.~(2).

Figure~2e and f respectively show the dependence of $P_y$ and $P_z$ on the orientation of the magnetic field, rotated in the $xy$ plane. In the two cases, the electric field was applied along the measured electric polarization component, i.e., along the $y$ or $z$ axis. Based on Eq.~(1) and the orthogonality of $\mathbf{L}$ to the magnetic field, the angular dependences of the two electric polarization components have the following forms:
\begin{eqnarray}
P_y \propto M_zL_x + M_xL_z= \eta \cos(\phi)\sin(\theta),\\
P_z \propto M_xL_y + M_yL_x= \eta \cos(2\phi)\cos(\theta).
\end{eqnarray}
where $\eta$ is a constant factor, $\phi$ is the angle between the magnetic field and the $x$ axis, and $\theta$ is the angle of $\mathbf{L}$ with respect to the $xy$ plane, as shown in Fig.~2b. Since the magnetic anisotropy is negligible, the angle $\theta$ is determined by the ME free energy. When the electric field is applied along the $y$ axis, the ME free energy is minimized by the $+L_z$ state ($\theta=+90\degree$) and the $-L_z$ state ($\theta=-90\degree$) for $0\degree<\phi<90\degree$  and $90\degree<\phi<180\degree$, respectively, as depicted in Fig.~2c (For the details of the calculation see the supplementary note 2 and supplementary figure 2). Thus, upon the rotation of the magnetic field, the $+L_z\rightarrow -L_z$ switching is expected when the magnetic and electric fields become parallel, while the back-switching should take place when they are antiparallel. The solid/dashed black curve in Fig.~2e indicates the calculated behaviour of $P_y$, according to Eq.~(3), for the $+L_z$/$-L_z$ mono-domain state. The angular dependence of the measured $P_y$ (purple curve in Fig.~2e) is well captured by this model, confirming the sudden switching between the $\pm L_z$ states.

When the electric field was applied along the $z$ axis, the ME energy is minimized by $\theta=0\degree$ for $-45\degree<\phi<45\degree$ and $135\degree<\phi<225\degree$, and by $\theta=180\degree$ for $45\degree<\phi<135\degree$ and $225\degree<\phi<315\degree$, as depicted in Fig. 2d. This means $\mathbf{L}$ lies within in the $xy$ plane during the entire rotation of $\mathbf{H}$. More specifically, the mutual orthogonality among $\mathbf{H}$, $\mathbf{E}$ and $\mathbf{L}$, according to Eq.~(1), results in the smooth rotation of $\mathbf{L}$, within each of the four $90\degree$ segments mentioned above. However, at the edges of these segments, i.e. at $\phi$ = $45\degree$, $135\degree$, $225\degree$ and $315\degree$, $\mathbf{L}$ is suddenly reversed. Therefore, if $\mathbf{L}$ precedes the magnetic field in one segment, then it follows the field in the next segment, as sketched in Fig.~2d. (For more details see supplementary note 2 and supplementary figure 3). The experimentally observed behaviour of $P_z$, as shown by purple curve in Fig.~2f, is also confirmed by our model. These results speak for the full controllability of $\mathbf{L}$ via the ME effect, and no significant influence of magnetic anisotropy on the kinetics of $\mathbf{L}$. Such an unconstrained in situ ME control of $\mathbf{L}$ even at $T\ll T_N$ makes Co$_3$O$_4$ unique among antiferromagnets. 

\vspace{0.2cm}

\textbf{Pulsed voltage control of AFM domains}

Next, we demonstrate the in situ switching of $\mathbf{L}$ by pulsed electric fields. Fig.~3a displays the effect of $+$, $+$, $-$, $-$, ... voltage pulse sequences on an initially multi-domain crystal at 4\,K. The pulsed electric field and the static magnetic field were oriented along the $z$ and $x$ axis, respectively. The current measured with the applied electric field exhibits peaks only when the sign of the voltage pulse is changed, i.e. at the first $+$ and first $-$ pulses. This current peak originates from the electric field induced switching of the ME polarization. The lack of current peaks associated with the second $+$ and the second $-$ pulses indicate that the first $+$ and $-$ pulses established the $+L_y$ and $-L_y$ mono-domain states, respectively.  During the switching process, the time dependence of $\alpha$ is obtained by dividing the electric polarization (the time-integral of the current) by the magnetic field value. The switching process is more easily accessible in Fig.~3b, where we directly plot the magnetoelectric coefficient versus the electric field, which resembles a typical ferroelectric hysteresis loop, and demonstrates that $\alpha$ is reversed when the electric field is inverted. Note that the AFM domain switching is non-volatile, as indicated by the high-remanent $\alpha$ observed after all electric pulses in Fig.~3b. This magnitude of remanent $\alpha$ agrees with the value in the inset of Fig.~1a, measured on a mono-domain sample after switching off the electric poling field. Though a similar in situ ME control of AFM domains has been reported in Cr$_2$O$_3$ and MnTiO$_3$, the switching in those cases are limited to the vicinity of $T_N$~\cite{he2010,iyama2013,kosub2017,sato2020}.

In Fig.~3c, we analyse the dependence of switching efficiency on the strength of the pulsed electric field in +14\,T at 4\,K. While electric fields $\gtrsim$ 9 kV/cm are required to create $\pm L_y$ mono-domain states, weaker electric pulses still lead to considerably high $\pm \alpha$ values in a reproducible fashion. This implies that a full control over $\mathbf{L}$ can be achieved by lower electric fields when microscopic AFM domains are subject to manipulation and not macroscopic volumes, like in the present case. A similar behaviour of $\alpha$ on the strength of the static magnetic field indicates that magnetic fields $\gtrsim$ 6\,T are necessary to create in situ $\pm L_y$ mono-domains by electric pulses of $\pm$17.4\,kV/cm, as seen in Fig.~3c.

Next, we investigate the dynamics of the switching with particular interest on the switching speed, a parameter highly relevant to device applications. The dependence of the switching time ($t_s$) on the rise time of the driving electric pulse ($t_r$) is shown in Fig.~3e. Time traces of the switching current are displayed in Fig.~3d for three different values of $t_r$, where $t_s$ is obtained as the full width of the current peak at half maximum. The switching is found to be as fast as $t_s\approx$ 5.5\,$\mu$s for $t_r=$ 10\,$\mu$s. We have to stress that for all $t_r$ values studied here, the switching was completed during the rise of the driving voltage, with the current peak located typically at $t\lesssim \frac{t_r}{2}$. This observation, together with the linear dependence of $t_s$ on $t_r$ without any sign of a saturation, suggests that the values of the switching time, indicated in Figs.~3d \& 3e, are limited by the rise time of the drive pulse of our experimental setup ($t_r$ $\geq$10\,$\mu$s), and the shortest intrinsic switching time is less than 5.5 $\mu$s. 

The sound velocity associated with the AFM magnon band in Co$_3$O$_4$, $v_m$ = 2.3\,km/s~\cite{zaharko2011}, imposes a fundamental limit for the speed of the AFM domain walls. This maximum domain wall speed leads to a switching time of $t_s$ = 100\,ps for 200\,nm sized AFM domain in Co$_3$O$_4$, which is considerably faster than $t_s$ = 100\,ns observed for heterostructure based on 200\,nm thick Cr$_2$O$_3$ films~\cite{toyoki2015}. Whether this fundamental speed limit can be reached or closely approached by AFM domain walls in Co$_3$O$_4$ requires additional studies, investigating the switching dynamics on the nano- to picosecond timescale. 

Fig.~3e also demonstrates that the change in $\alpha$ ($\Delta\alpha$) associated with the switching between $\pm L_y$, i.e. the switching efficiency, remains unchanged while the rise time of the driving pulse is reduced to $t_r$ = 10 $\mu$s. Finally, we show that $\mathbf{L}$ can be switched reversibly and repetitively in Co$_3$O$_4$. The switching efficiency is not reduced upon subsequent switchings of $\mathbf{L}$ up to million switching events, as seen in Fig.~3f. Instead a weak increase is observed, which is attributed to the training effect, often observed in ferroelectrics~\cite{gao2007,glaum2012}. 

\vspace{0.2cm}


In summary, we demonstrated the fast manipulation of AFM states in the isotropic collinear antiferromagnet Co$_3$O$_4$ by utilizing its strong linear ME effect. We showed that the AFM N\'eel vector can be either rotated smoothly or reversed isothermally by electric fields. Furthermore, our results revealed the microsecond dynamics of the non-volatile and reversible switching process of macroscopic AFM domains and indicate that the switching process can be even faster for microscopic domains. The present study represents an important step towards the on-demand electric control of AFM states, that may eventually lead to the realization of fast-operation AFM spintronic device.

\textbf{Data availability.}~The experimental data that support the findings of this manuscript are available from the corresponding author upon reasonable request.

\bibliography{Co3O4_voltage_AFM_Domains}

\begin{thebibliography}{10}
\expandafter\ifx\csname url\endcsname\relax
  \def\url#1{\texttt{#1}}\fi
\expandafter\ifx\csname urlprefix\endcsname\relax\def\urlprefix{URL }\fi
\providecommand{\bibinfo}[2]{#2}
\providecommand{\eprint}[2][]{\url{#2}}

\bibitem{vzutic2004}
\bibinfo{author}{{\v{Z}}uti{\'c}, I.}, \bibinfo{author}{Fabian, J.} \&
  \bibinfo{author}{Sarma, S.~D.}
\newblock \bibinfo{title}{Spintronics: Fundamentals and applications}.
\newblock \emph{\bibinfo{journal}{Rev. Mod. Phys.}}
  \textbf{\bibinfo{volume}{76}}, \bibinfo{pages}{323} (\bibinfo{year}{2004}).

\bibitem{dietl2014}
\bibinfo{author}{Dietl, T.} \& \bibinfo{author}{Ohno, H.}
\newblock \bibinfo{title}{Dilute ferromagnetic semiconductors: Physics and
  spintronic structures}.
\newblock \emph{\bibinfo{journal}{Rev. Mod. Phys.}}
  \textbf{\bibinfo{volume}{86}}, \bibinfo{pages}{187} (\bibinfo{year}{2014}).

\bibitem{jungwirth2016}
\bibinfo{author}{Jungwirth, T.}, \bibinfo{author}{Marti, X.},
  \bibinfo{author}{Wadley, P.} \& \bibinfo{author}{Wunderlich, J.}
\newblock \bibinfo{title}{Antiferromagnetic spintronics}.
\newblock \emph{\bibinfo{journal}{Nat. Nanotechnol.}}
  \textbf{\bibinfo{volume}{11}}, \bibinfo{pages}{231--241}
  (\bibinfo{year}{2016}).

\bibitem{baltz2018}
\bibinfo{author}{Baltz, V.} \emph{et~al.}
\newblock \bibinfo{title}{Antiferromagnetic spintronics}.
\newblock \emph{\bibinfo{journal}{Rev. Mod. Phys.}}
  \textbf{\bibinfo{volume}{90}}, \bibinfo{pages}{015005}
  (\bibinfo{year}{2018}).

\bibitem{neel1971}
\bibinfo{author}{N{\'e}el, L.}
\newblock \bibinfo{title}{Magnetism and local molecular field}.
\newblock \emph{\bibinfo{journal}{Science}} \textbf{\bibinfo{volume}{174}},
  \bibinfo{pages}{985--992} (\bibinfo{year}{1971}).

\bibitem{jungwirth2018natphys}
\bibinfo{author}{Jungwirth, T.} \emph{et~al.}
\newblock \bibinfo{title}{The multiple directions of antiferromagnetic
  spintronics}.
\newblock \emph{\bibinfo{journal}{Nat. Phys.}} \textbf{\bibinfo{volume}{14}},
  \bibinfo{pages}{200--203} (\bibinfo{year}{2018}).

\bibitem{jungfleisch2018}
\bibinfo{author}{Jungfleisch, M.~B.}, \bibinfo{author}{Zhang, W.} \&
  \bibinfo{author}{Hoffmann, A.}
\newblock \bibinfo{title}{Perspectives of antiferromagnetic spintronics}.
\newblock \emph{\bibinfo{journal}{Phys. Lett. A}}
  \textbf{\bibinfo{volume}{382}}, \bibinfo{pages}{865--871}
  (\bibinfo{year}{2018}).

\bibitem{fina2014}
\bibinfo{author}{Fina, I.} \emph{et~al.}
\newblock \bibinfo{title}{Anisotropic magnetoresistance in an antiferromagnetic
  semiconductor}.
\newblock \emph{\bibinfo{journal}{Nat. Commun.}} \textbf{\bibinfo{volume}{5}},
  \bibinfo{pages}{1--7} (\bibinfo{year}{2014}).

\bibitem{wang2014}
\bibinfo{author}{Wang, C.} \emph{et~al.}
\newblock \bibinfo{title}{Anisotropic magnetoresistance in antiferromagnetic
  {Sr$_2$IrO$_4$}}.
\newblock \emph{\bibinfo{journal}{Phys. Rev. X}} \textbf{\bibinfo{volume}{4}},
  \bibinfo{pages}{041034} (\bibinfo{year}{2014}).

\bibitem{zhang2014}
\bibinfo{author}{Zhang, W.} \emph{et~al.}
\newblock \bibinfo{title}{Spin hall effects in metallic antiferromagnets}.
\newblock \emph{\bibinfo{journal}{Phys. Rev. Lett.}}
  \textbf{\bibinfo{volume}{113}}, \bibinfo{pages}{196602}
  (\bibinfo{year}{2014}).

\bibitem{kimata2019}
\bibinfo{author}{Kimata, M.} \emph{et~al.}
\newblock \bibinfo{title}{Magnetic and magnetic inverse spin hall effects in a
  non-collinear antiferromagnet}.
\newblock \emph{\bibinfo{journal}{Nature}} \textbf{\bibinfo{volume}{565}},
  \bibinfo{pages}{627--630} (\bibinfo{year}{2019}).

\bibitem{chen2021}
\bibinfo{author}{Chen, X.} \emph{et~al.}
\newblock \bibinfo{title}{Observation of the antiferromagnetic spin hall
  effect}.
\newblock \emph{\bibinfo{journal}{Nat. Mater.}} \bibinfo{pages}{1--5}
  (\bibinfo{year}{2021}).

\bibitem{seki2015}
\bibinfo{author}{Seki, S.} \emph{et~al.}
\newblock \bibinfo{title}{Thermal generation of spin current in an
  antiferromagnet}.
\newblock \emph{\bibinfo{journal}{Phys. Rev. Lett.}}
  \textbf{\bibinfo{volume}{115}}, \bibinfo{pages}{266601}
  (\bibinfo{year}{2015}).

\bibitem{wu2016}
\bibinfo{author}{Wu, S.~M.} \emph{et~al.}
\newblock \bibinfo{title}{Antiferromagnetic spin seebeck effect}.
\newblock \emph{\bibinfo{journal}{Phys. Rev. Lett.}}
  \textbf{\bibinfo{volume}{116}}, \bibinfo{pages}{097204}
  (\bibinfo{year}{2016}).

\bibitem{rezende2016}
\bibinfo{author}{Rezende, S.}, \bibinfo{author}{Rodr{\'\i}guez-Su{\'a}rez, R.}
  \& \bibinfo{author}{Azevedo, A.}
\newblock \bibinfo{title}{Theory of the spin seebeck effect in
  antiferromagnets}.
\newblock \emph{\bibinfo{journal}{Phys. Rev. B}} \textbf{\bibinfo{volume}{93}},
  \bibinfo{pages}{014425} (\bibinfo{year}{2016}).

\bibitem{zelezny2014}
\bibinfo{author}{{\v{Z}}elezn{\`y}, J.} \emph{et~al.}
\newblock \bibinfo{title}{Relativistic {N{\'e}el}-order fields induced by
  electrical current in antiferromagnets}.
\newblock \emph{\bibinfo{journal}{Phys. Rev. Lett.}}
  \textbf{\bibinfo{volume}{113}}, \bibinfo{pages}{157201}
  (\bibinfo{year}{2014}).

\bibitem{wadley2016}
\bibinfo{author}{Wadley, P.} \emph{et~al.}
\newblock \bibinfo{title}{Electrical switching of an antiferromagnet}.
\newblock \emph{\bibinfo{journal}{Science}} \textbf{\bibinfo{volume}{351}},
  \bibinfo{pages}{587--590} (\bibinfo{year}{2016}).

\bibitem{kampfrath2011}
\bibinfo{author}{Kampfrath, T.} \emph{et~al.}
\newblock \bibinfo{title}{Coherent terahertz control of antiferromagnetic spin
  waves}.
\newblock \emph{\bibinfo{journal}{Nat. Photonics}}
  \textbf{\bibinfo{volume}{5}}, \bibinfo{pages}{31--34} (\bibinfo{year}{2011}).

\bibitem{gao2020}
\bibinfo{author}{Gao, S.} \emph{et~al.}
\newblock \bibinfo{title}{Fractional antiferromagnetic skyrmion lattice induced
  by anisotropic couplings}.
\newblock \emph{\bibinfo{journal}{Nature}} \bibinfo{pages}{1--5}
  (\bibinfo{year}{2020}).

\bibitem{legrand2020}
\bibinfo{author}{Legrand, W.} \emph{et~al.}
\newblock \bibinfo{title}{Room-temperature stabilization of antiferromagnetic
  skyrmions in synthetic antiferromagnets}.
\newblock \emph{\bibinfo{journal}{Nat. Mater.}} \textbf{\bibinfo{volume}{19}},
  \bibinfo{pages}{34--42} (\bibinfo{year}{2020}).

\bibitem{zhang2016}
\bibinfo{author}{Zhang, X.}, \bibinfo{author}{Zhou, Y.} \&
  \bibinfo{author}{Ezawa, M.}
\newblock \bibinfo{title}{Magnetic bilayer-skyrmions without skyrmion hall
  effect}.
\newblock \emph{\bibinfo{journal}{Nat. Commun.}} \textbf{\bibinfo{volume}{7}},
  \bibinfo{pages}{1--7} (\bibinfo{year}{2016}).

\bibitem{woo2018}
\bibinfo{author}{Woo, S.} \emph{et~al.}
\newblock \bibinfo{title}{Current-driven dynamics and inhibition of the
  skyrmion hall effect of ferrimagnetic skyrmions in {GdFeCo} films}.
\newblock \emph{\bibinfo{journal}{Nat. Commun.}} \textbf{\bibinfo{volume}{9}},
  \bibinfo{pages}{1--8} (\bibinfo{year}{2018}).

\bibitem{duine2018}
\bibinfo{author}{Duine, R.}, \bibinfo{author}{Lee, K.-J.},
  \bibinfo{author}{Parkin, S.~S.} \& \bibinfo{author}{Stiles, M.~D.}
\newblock \bibinfo{title}{Synthetic antiferromagnetic spintronics}.
\newblock \emph{\bibinfo{journal}{Nat. Phys.}} \textbf{\bibinfo{volume}{14}},
  \bibinfo{pages}{217--219} (\bibinfo{year}{2018}).

\bibitem{song2018}
\bibinfo{author}{Song, C.} \emph{et~al.}
\newblock \bibinfo{title}{How to manipulate magnetic states of
  antiferromagnets}.
\newblock \emph{\bibinfo{journal}{Nanotechnology}}
  \textbf{\bibinfo{volume}{29}}, \bibinfo{pages}{112001}
  (\bibinfo{year}{2018}).

\bibitem{leo2018}
\bibinfo{author}{Leo, N.} \emph{et~al.}
\newblock \bibinfo{title}{Magnetoelectric inversion of domain patterns}.
\newblock \emph{\bibinfo{journal}{Nature}} \textbf{\bibinfo{volume}{560}},
  \bibinfo{pages}{466--470} (\bibinfo{year}{2018}).

\bibitem{parthasarathy2019}
\bibinfo{author}{Parthasarathy, A.} \& \bibinfo{author}{Rakheja, S.}
\newblock \bibinfo{title}{Dynamics of magnetoelectric reversal of an
  antiferromagnetic domain}.
\newblock \emph{\bibinfo{journal}{Phys. Rev. Appl.}}
  \textbf{\bibinfo{volume}{11}}, \bibinfo{pages}{034051}
  (\bibinfo{year}{2019}).

\bibitem{baldrati2020}
\bibinfo{author}{Baldrati, L.} \emph{et~al.}
\newblock \bibinfo{title}{Efficient spin torques in antiferromagnetic {CoO/Pt}
  quantified by comparing field-and current-induced switching}.
\newblock \emph{\bibinfo{journal}{Phys. Rev. Lett.}}
  \textbf{\bibinfo{volume}{125}}, \bibinfo{pages}{077201}
  (\bibinfo{year}{2020}).

\bibitem{yan2020}
\bibinfo{author}{Yan, H.} \emph{et~al.}
\newblock \bibinfo{title}{Electric-field-controlled antiferromagnetic
  spintronic devices}.
\newblock \emph{\bibinfo{journal}{Adv. Mater.}} \textbf{\bibinfo{volume}{32}},
  \bibinfo{pages}{1905603} (\bibinfo{year}{2020}).

\bibitem{thole2020}
\bibinfo{author}{Th{\"o}le, F.}, \bibinfo{author}{Keliri, A.} \&
  \bibinfo{author}{Spaldin, N.~A.}
\newblock \bibinfo{title}{Concepts from the linear magnetoelectric effect that
  might be useful for antiferromagnetic spintronics}.
\newblock \emph{\bibinfo{journal}{J. Appl. Phys.}}
  \textbf{\bibinfo{volume}{127}}, \bibinfo{pages}{213905}
  (\bibinfo{year}{2020}).

\bibitem{bodnar2018}
\bibinfo{author}{Bodnar, S.~Y.} \emph{et~al.}
\newblock \bibinfo{title}{Writing and reading antiferromagnetic {Mn$_2$Au} by
  {N{\'e}el} spin-orbit torques and large anisotropic magnetoresistance}.
\newblock \emph{\bibinfo{journal}{Nat. Commun.}} \textbf{\bibinfo{volume}{9}},
  \bibinfo{pages}{1--7} (\bibinfo{year}{2018}).

\bibitem{nair2020}
\bibinfo{author}{Nair, N.~L.} \emph{et~al.}
\newblock \bibinfo{title}{Electrical switching in a magnetically intercalated
  transition metal dichalcogenide}.
\newblock \emph{\bibinfo{journal}{Nat. Mater.}} \textbf{\bibinfo{volume}{19}},
  \bibinfo{pages}{153--157} (\bibinfo{year}{2020}).

\bibitem{borisov2005}
\bibinfo{author}{Borisov, P.}, \bibinfo{author}{Hochstrat, A.},
  \bibinfo{author}{Chen, X.}, \bibinfo{author}{Kleemann, W.} \&
  \bibinfo{author}{Binek, C.}
\newblock \bibinfo{title}{Magnetoelectric switching of exchange bias}.
\newblock \emph{\bibinfo{journal}{Phys. Rev. Lett.}}
  \textbf{\bibinfo{volume}{94}}, \bibinfo{pages}{117203}
  (\bibinfo{year}{2005}).

\bibitem{iyama2013}
\bibinfo{author}{Iyama, A.} \& \bibinfo{author}{Kimura, T.}
\newblock \bibinfo{title}{Magnetoelectric hysteresis loops in {Cr$_2$O$_3$} at
  room temperature}.
\newblock \emph{\bibinfo{journal}{Phys. Rev. B}} \textbf{\bibinfo{volume}{87}},
  \bibinfo{pages}{180408} (\bibinfo{year}{2013}).

\bibitem{kocsis2018}
\bibinfo{author}{Kocsis, V.} \emph{et~al.}
\newblock \bibinfo{title}{Identification of antiferromagnetic domains via the
  optical magnetoelectric effect}.
\newblock \emph{\bibinfo{journal}{Phys. Rev. Lett.}}
  \textbf{\bibinfo{volume}{121}}, \bibinfo{pages}{057601}
  (\bibinfo{year}{2018}).

\bibitem{van2007}
\bibinfo{author}{Van~Aken, B.~B.}, \bibinfo{author}{Rivera, J.-P.},
  \bibinfo{author}{Schmid, H.} \& \bibinfo{author}{Fiebig, M.}
\newblock \bibinfo{title}{Observation of ferrotoroidic domains}.
\newblock \emph{\bibinfo{journal}{Nature}} \textbf{\bibinfo{volume}{449}},
  \bibinfo{pages}{702--705} (\bibinfo{year}{2007}).

\bibitem{zimmermann2014}
\bibinfo{author}{Zimmermann, A.~S.}, \bibinfo{author}{Meier, D.} \&
  \bibinfo{author}{Fiebig, M.}
\newblock \bibinfo{title}{Ferroic nature of magnetic toroidal order}.
\newblock \emph{\bibinfo{journal}{Nat. Commun.}} \textbf{\bibinfo{volume}{5}},
  \bibinfo{pages}{1--6} (\bibinfo{year}{2014}).

\bibitem{sato2020}
\bibinfo{author}{Sato, T.}, \bibinfo{author}{Abe, N.}, \bibinfo{author}{Kimura,
  S.}, \bibinfo{author}{Tokunaga, Y.} \& \bibinfo{author}{Arima, T.-h.}
\newblock \bibinfo{title}{Magnetochiral dichroism in a collinear
  antiferromagnet with no magnetization}.
\newblock \emph{\bibinfo{journal}{Phys. Rev. Lett.}}
  \textbf{\bibinfo{volume}{124}}, \bibinfo{pages}{217402}
  (\bibinfo{year}{2020}).

\bibitem{mufti2011}
\bibinfo{author}{Mufti, N.} \emph{et~al.}
\newblock \bibinfo{title}{Magnetoelectric coupling in {MnTiO$_3$}}.
\newblock \emph{\bibinfo{journal}{Phys. Rev. B}} \textbf{\bibinfo{volume}{83}},
  \bibinfo{pages}{104416} (\bibinfo{year}{2011}).

\bibitem{kosub2017}
\bibinfo{author}{Kosub, T.} \emph{et~al.}
\newblock \bibinfo{title}{Purely antiferromagnetic magnetoelectric random
  access memory}.
\newblock \emph{\bibinfo{journal}{Nat. Commun.}} \textbf{\bibinfo{volume}{8}},
  \bibinfo{pages}{1--7} (\bibinfo{year}{2017}).

\bibitem{he2010}
\bibinfo{author}{He, X.} \emph{et~al.}
\newblock \bibinfo{title}{Robust isothermal electric control of exchange bias
  at room temperature}.
\newblock \emph{\bibinfo{journal}{Nat. Mater.}} \textbf{\bibinfo{volume}{9}},
  \bibinfo{pages}{579--585} (\bibinfo{year}{2010}).

\bibitem{roth1964}
\bibinfo{author}{Roth, W.}
\newblock \bibinfo{title}{The magnetic structure of {Co$_3$O$_4$}}.
\newblock \emph{\bibinfo{journal}{J. Phys. Chem. Solids}}
  \textbf{\bibinfo{volume}{25}}, \bibinfo{pages}{1--10} (\bibinfo{year}{1964}).

\bibitem{cova2019}
\bibinfo{author}{Cova, F.}, \bibinfo{author}{Blanco, M.~V.},
  \bibinfo{author}{Hanfland, M.} \& \bibinfo{author}{Garbarino, G.}
\newblock \bibinfo{title}{Study of the high pressure phase evolution of
  {Co$_3$O$_4$}}.
\newblock \emph{\bibinfo{journal}{Phys. Rev. B}}
  \textbf{\bibinfo{volume}{100}}, \bibinfo{pages}{054111}
  (\bibinfo{year}{2019}).

\bibitem{sunda2021}
\bibinfo{author}{Sundaresan, A.} \& \bibinfo{author}{Ter-Oganessian, N.}
\newblock \bibinfo{title}{Magnetoelectric and multiferroic properties of
  spinels}.
\newblock \emph{\bibinfo{journal}{J. Appl. Phys.}}
  \textbf{\bibinfo{volume}{129}}, \bibinfo{pages}{060901}
  (\bibinfo{year}{2021}).

\bibitem{saha2016}
\bibinfo{author}{Saha, R.} \emph{et~al.}
\newblock \bibinfo{title}{Magnetoelectric effect in simple collinear
  antiferromagnetic spinels}.
\newblock \emph{\bibinfo{journal}{Phys. Rev. B}} \textbf{\bibinfo{volume}{94}},
  \bibinfo{pages}{014428} (\bibinfo{year}{2016}).

\bibitem{gao2017}
\bibinfo{author}{Gao, S.} \emph{et~al.}
\newblock \bibinfo{title}{Spiral spin-liquid and the emergence of a vortex-like
  state in {MnSc$_2$S$_4$}}.
\newblock \emph{\bibinfo{journal}{Nat. Phys.}} \textbf{\bibinfo{volume}{13}},
  \bibinfo{pages}{157--161} (\bibinfo{year}{2017}).

\bibitem{fritsch2004}
\bibinfo{author}{Fritsch, V.} \emph{et~al.}
\newblock \bibinfo{title}{Spin and orbital frustration in {MnSc$_2$S$_4$} and
  {FeSc$_2$S$_4$}}.
\newblock \emph{\bibinfo{journal}{Phys. Rev. Lett.}}
  \textbf{\bibinfo{volume}{92}}, \bibinfo{pages}{116401}
  (\bibinfo{year}{2004}).

\bibitem{ter2014}
\bibinfo{author}{Ter-Oganessian, N.}
\newblock \bibinfo{title}{Cation-ordered {A$_{1/2}$A$'$$_{1/2}$B$_2$X$_4$}
  magnetic spinels as magnetoelectrics}.
\newblock \emph{\bibinfo{journal}{J. Magn. Magn. Mater.}}
  \textbf{\bibinfo{volume}{364}}, \bibinfo{pages}{47--54}
  (\bibinfo{year}{2014}).

\bibitem{ghara2017}
\bibinfo{author}{Ghara, S.}, \bibinfo{author}{Ter-Oganessian, N.} \&
  \bibinfo{author}{Sundaresan, A.}
\newblock \bibinfo{title}{Linear magnetoelectric effect as a signature of
  long-range collinear antiferromagnetic ordering in the frustrated spinel
  {CoAl$_2$O$_4$}}.
\newblock \emph{\bibinfo{journal}{Phys. Rev. B}} \textbf{\bibinfo{volume}{95}},
  \bibinfo{pages}{094404} (\bibinfo{year}{2017}).

\bibitem{de2021}
\bibinfo{author}{De, C.} \emph{et~al.}
\newblock \bibinfo{title}{Magnetoelectric effect in a single crystal of the
  frustrated spinel {CoAl$_2$O$_4$}}.
\newblock \emph{\bibinfo{journal}{Phys. Rev. B}}
  \textbf{\bibinfo{volume}{103}}, \bibinfo{pages}{094406}
  (\bibinfo{year}{2021}).

\bibitem{zaharko2011}
\bibinfo{author}{Zaharko, O.} \emph{et~al.}
\newblock \bibinfo{title}{Spin liquid in a single crystal of the frustrated
  diamond lattice antiferromagnet {CoAl$_2$O$_4$}}.
\newblock \emph{\bibinfo{journal}{Phys. Rev. B}} \textbf{\bibinfo{volume}{84}},
  \bibinfo{pages}{094403} (\bibinfo{year}{2011}).

\bibitem{toyoki2015}
\bibinfo{author}{Toyoki, K.} \emph{et~al.}
\newblock \bibinfo{title}{Magnetoelectric switching of perpendicular exchange
  bias in {Pt}/{Co}/$\alpha$-{Cr$_2$O$_3$}/{Pt} stacked films}.
\newblock \emph{\bibinfo{journal}{Appl. Phys. Lett.}}
  \textbf{\bibinfo{volume}{106}}, \bibinfo{pages}{162404}
  (\bibinfo{year}{2015}).

\bibitem{gao2007}
\bibinfo{author}{Gao, Y.}, \bibinfo{author}{Uchino, K.} \&
  \bibinfo{author}{Viehland, D.}
\newblock \bibinfo{title}{Domain wall release in “hard” piezoelectric under
  continuous large amplitude ac excitation}.
\newblock \emph{\bibinfo{journal}{J. Appl. Phys.}}
  \textbf{\bibinfo{volume}{101}}, \bibinfo{pages}{114110}
  (\bibinfo{year}{2007}).

\bibitem{glaum2012}
\bibinfo{author}{Glaum, J.}, \bibinfo{author}{Genenko, Y.~A.},
  \bibinfo{author}{Kungl, H.}, \bibinfo{author}{Ana~Schmitt, L.} \&
  \bibinfo{author}{Granzow, T.}
\newblock \bibinfo{title}{De-aging of {Fe}-doped lead-zirconate-titanate
  ceramics by electric field cycling: 180$\degree$-vs. non-180$\degree$ domain
  wall processes}.
\newblock \emph{\bibinfo{journal}{J. Appl. Phys.}}
  \textbf{\bibinfo{volume}{112}}, \bibinfo{pages}{034103}
  (\bibinfo{year}{2012}).

\end{thebibliography}

\section*{Methods}
\textbf{Sample synthesis.} The single crystals of Co$_3$O$_4$ were grown by chemical transport reaction method using the preliminary treated high purity powder of Co$_3$O$_4$ together with anhydrous TeCl$_4$ transport agent. The growth was performed in a horizontal two-zone furnace with the source temperature of 830 $\degree$C and a temperature gradient of 40 to 25\,$\degree$C. The crystallographic orientation of the crystals were determined by X-ray Laue diffraction.

\textbf{ME measurements.} Crystals with flat (001) surfaces were prepared for the measurements. Contacts on two parallel (001) surfaces were made by silver paint. All measurements were carried out in an Oxford helium-flow cryostat, where the electrical measurements were performed up to the lowest temperature of 4\,K and the highest magnetic field of 14\,T. Temperature- and magnetic field-dependent electric polarization were obtained via the pyroelectric and ME current measurements, respectively, with a Keysight electrometer (B2987A). For angular-dependent electric polarization study, the ME current was recorded under a rotating magnetic field of 14\,T at a constant electric field of +17.4\,kV/cm by using a home-designed probe, where the sample platform can be rotated by a stepper motor. For the data shown in Fig.~1, the sample was cooled with $E_p$ and $H_p$ from 45\,K to 4\,K and the pyroelectric/ME current was measured after switching off the electric field. For the data shown in Fig.~2 \& 3, the measurements were performed on a virgin multi-domain sample without any ME cooling. In all these cases, the electric polarization were obtained by performing integration of the measured pyroelectric/ME current over time.

\textbf{Pulsed electric field measurements.} Electric polarization measurements in pulsed electric fields were carried out by a AixACCT TF Analyzer 2000 equipped with high voltage booster HVB 1000 and Krohn-Hite model 7500 amplifier. In order to obtain the purely ME contribution, we subtracted the current arising from the dielectric response in zero magnetic field from the current measured in applied magnetic fields. The data indicated by the closed and opened symbols in Fig.~3e were obtained by the pulse field measurements on crystal with thickness of 250 $\mu$m and 40 $\mu$m, respectively.

\textbf{Acknowledgements}~We thank Anton Jesche for assistance in X-ray Laue diffraction experiment and Peter Lunkenheimer, S\'andor Bord\'acs, Maxim Mostovoy, N. V. Ter-Oganessian and Sundaresan Athinarayanan for fruitful discussions. This work was supported by the DFG via the Transregional Research Collaboration TRR 80 From Electronic Correlations to Functionality (Augsburg/Munich/Stuttgart).

\textbf{Author Contributions:}~S.G. initiated this project; L.P. and V.T. synthesized the single crystals; M.W., S.G., K.G. and S.K. performed the experiments and analyzed the data; S.G. and I.K. performed the theoretical modeling and wrote the manuscript; I.K. supervised this project. All authors discussed the content of the manuscript.

\textbf{Competing interests:}~The authors declare no competing interests.

\end{document}